\documentclass[%
 reprint,
superscriptaddress,
 amsmath,amssymb,
aps,
prb,
]{revtex4-2}
\usepackage{bm,latexsym,mathrsfs,enumerate}
\usepackage{graphicx}
\usepackage{dcolumn}
\usepackage{bm}
\usepackage{hyperref}
\usepackage[mathlines]{lineno}
\usepackage{xcolor}
\usepackage{hyperref}

\begin{document}

\title{Morphology of Polarization States in Strained Ferroelectric Films}

\author{L{\'e}o Boron}
\affiliation{Laboratoire de Physique de la Mati{\`e}re Condens{\'e}e, Universit{\'e} de Picardie Jules Verne, 33 rue Saint Leu, Amiens 80039, France}

\author{Anaïs Sen{\'e}}
\affiliation{Laboratoire de Physique de la Mati{\`e}re Condens{\'e}e, Universit{\'e} de Picardie Jules Verne, 33 rue Saint Leu, Amiens 80039, France}

\author{Yuri Tikhonov}
\affiliation{Laboratoire de Physique de la Mati{\`e}re Condens{\'e}e, Universit{\'e} de Picardie Jules Verne, 33 rue Saint Leu, Amiens 80039, France}

\author{Anna Razumnaya}%
\email{anna.razumnaya@ijs.si}
\affiliation{Jozef Stefan Institute, Jamova Cesta 39, Ljubljana 1000, Slovenia}

\author{Igor Lukyanchuk}
\affiliation{Laboratoire de Physique de la Mati{\`e}re Condens{\'e}e, Universit{\'e} de Picardie Jules Verne, 33 rue Saint Leu, Amiens 80039, France}

\author{Svitlana Kondovych}
\email{s.kondovych@ifw-dresden.de}
\affiliation{Institute for Theoretical Solid State Physics, Leibniz Institute for Solid State and Materials Research Dresden, Helmholzstr. 20. Dresden D-01069, Germany}


\begin{abstract}
Ferroelectric thin films under epitaxial strain exhibit a variety of vortex-like topological polarization textures. To analyze them, we build on the Ginzburg-Landau-Devonshire framework and extend the previously introduced soft-domain approach. This formulation provides a compact variational theoretical description of polarization morphologies in strained PbTiO$_3$ films. It yields phase diagrams as a function of temperature, strain, and thickness, and clarifies the morphological structure of emergent topological states. The method is computationally efficient and offers practical guidance for experimental studies of ferroelectric nanostructures.
\end{abstract}


\maketitle

\section{Introduction}

Ferroelectric thin films under epitaxial strain have recently emerged as a rich platform for observing complex polarization textures that go far beyond classical domain patterns~\cite{Das2020,Junquera2023,Lukyanchuk2024,Lukyanchuk2025}. These systems often stabilize states where the polarization rotates smoothly in space and organizes into swirled topological configurations. Such structures form to reduce the depolarizing energy that arises in uniformly polarized films when polarization terminates at the surfaces, creating bound charges.

Advanced microscopy techniques have revealed a variety of such states, including ordered vortex-antivortex arrays, flux-closure patterns, and more intricate textures such as helices, bubbles,  and labyrinthine modulations~\cite{Yadav2016,Yadav2019,Shafer2018,Das2019,Wang2020,Gong2021,Behera2022,Lichtensteiger2014,Zhang2017,Nahas2020,Nahas2020b}. These findings underline that strain, film thickness, and electrostatics jointly create a landscape where topological polarization states naturally arise.

These textures are well reproduced by atomistic simulations and phase-field modeling approaches~\cite{Naumov2004,Schlom2007,Schlom2008,Stachiotti2011,Yadav2016,Das2019,Nahas2020}, which capture their microscopic structure with high fidelity. However, since these approaches are computationally intensive and typically focus on local configurations or specific geometries, achieving a global, systematic overview of polarization organization and phase stability remains challenging.

Theoretical approaches have long provided a useful description of ferroelectric instabilities, strain effects, and domain formation. Earlier studies mostly addressed these issues at a basic level, focusing either on strain-tunable uniform states~\cite{Pertsev1998,Pertsev1999} or on the classical Landau-Kittel mechanism of polarization domain formation~\cite{Bratkovsky2000}. Using the Ginzburg-Landau-Devonshire (GLD) framework, more recent studies have calculated the polarization profiles within domain structures, introducing the concept of soft domains characterized by a smooth variation of polarization both across the film thickness and through the domain walls~\cite{Stephanovich2003,Stephanovich2005,Deguerville2005,Baudry2015,Lukyanchuk2009}. This type of smooth polarization distribution can be regarded as a precursor stage that naturally anticipates the more complex vortex-like and topological states later revealed experimentally. 

Building on this perspective, the soft-domain concept opened the way toward analyzing continuous polarization textures and provided a natural link to the description of complex polarization states that have been reported over the past decade.
 In the present work, we further develop this framework into a systematic analysis of topological polarization morphologies in strained PbTiO$_3$ films. A key advantage of the method is that it enables the quantitative evaluation of the stability of different configurations and the construction of phase diagrams, while avoiding the high computational cost associated with large-scale atomistic or phase-field simulations. Our approach develops a morphological classification of vortex-, helix-, wave-, and combined-type states, and identifies the conditions under which they replace uniform or classical domain configurations.

The paper is organized as follows. In Sec.~\ref{SecBasicEquations} we present the GLD functional and its coupling to elastic strain and electric fields. Section~\ref{SecSoftDomains} introduces the soft-domain approach and develops the effective free-energy description of vortex-type states. In Sec.~\ref{SecPolarizationStates} we construct the corresponding phase diagrams and discuss them in relation to phase-field modeling results.
Section~\ref{SecChallenges} compares the analytical results with experimental observations and outlines the main challenges of the soft-domain approach.
 The conclusion, summarizing the main results and outlining future directions, is given in Sec.~\ref{SecConclusion}.

\section{Basic equations}
\label{SecBasicEquations}
\subsection{Variational functional}

The theoretical framework of this work is the GLD approach, which provides a phenomenological description of ferroelectric systems in the vicinity of the Curie temperature $T_c$. The order parameter is the macroscopic polarization field $\mathbf{P}$, coupled to the elastic strain tensor $u_{ij}$ and the electric field $\mathbf{E}$. The total free energy functional is given by
\begin{equation}
\mathcal{F} = \int_V F\, dV ,
\label{FunctGLD}
\end{equation}
where the free energy density $F$ consists of the uniform contribution $F_{P}$, the gradient contribution $F_{\text{grad}}$, the elastic contribution $F_{u}$, and the electrostatic contribution $F_{E}$, which are discussed in the following sections.
 
\subsection{Uniform energy}
The uniform energy density $F_{P}$ is written as a polynomial expansion in the polarization components $\mathbf{P}=(P_1,P_2,P_3)$,

\begin{equation}
F_P =\left[a_{i}\left( T\right)
P_{i}^{2}+a_{ij}P_{i}^{2}P_{j}^{2}+a_{ijk}P_{i}^{2}P_{j}^{2}P_{k}^{2}
\right] _{i\leq j\leq k} ,
\label{FP}    
\end{equation}
where the summation convention over spatial indices $i,j,k \in \{1,2,3\}$ (or $\{x,y,z\}$) is implied. 

The temperature-dependent coefficients $a{_i}(T)$ of the quadratic terms become negative below $T_c$, providing the instability of the paraelectric phase toward the emergence of a spontaneous polarization. The stability of the ferroelectric phase is then ensured by the fourth- and sixth-order terms, which determine the magnitude of $\mathbf{P}$ and select the equilibrium directions of polarization.  

The explicit form of the expansion coefficients depends on the underlying crystal symmetry.
For a cubic parent phase, all Cartesian axes are equivalent and the coefficients are invariant under arbitrary permutations of indices $\{1,2,3\}$.
In the tetragonal case, the $x$- and $y$-axes remain equivalent ($1 \leftrightarrow 2$), while the $z$-axis ($3$) becomes distinct, leading to a partial splitting of the coefficients.
In the rhombohedral case, no permutation symmetry among Cartesian indices is preserved, and the coefficients along different directions are generally nonequivalent.

\subsection{Gradient energy}
The gradient contribution $F_{\text{grad}}$ describes the energy cost associated with spatial inhomogeneities of the polarization field. In tensorial notation it reads
\begin{equation}
F_{\mathrm{grad}}=\tfrac{1}{2}\,G_{ijkl}\,(\partial_i P_{j})( \partial_k P_{l}),
\label{Fgrad}
\end{equation}
where $G_{ijkl}$ are the gradient stiffness coefficients constrained by crystal symmetry.
For PbTiO$_3$, the reported values~\cite{Wang2004} (in units of $10^{10}\,\mathrm{C^{-2}m^{4}N}$) are $G_{1111}=2.77$, $G_{1122}=0$, $G_{1212}=1.38$, together with all equivalent permutations of indices dictated by cubic symmetry.

It is convenient to rewrite the gradient energy in the Frank-Oseen-type representation, originally introduced for nematic liquid crystals~\cite{DeGennes1993}:
\begin{gather}
   F_{\mathrm{grad}}
   =\frac{1}{2}K_{1}(\nabla\!\cdot\!\mathbf{P})^{2}
   +\frac{1}{2}K_{2}(\nabla\!\times\!\mathbf{P})^{2} \notag \\
   +\frac{1}{2}K_{\mathrm{a}}\!\left[(\partial_{x}P_{1})^{2}
   +(\partial_{y}P_{2})^{2}
   +(\partial_{z}P_{3})^{2}\right],
\label{Frank}
\end{gather}
where $K_{1}$, $K_{2}$, and $K_{\mathrm{a}}$ denote the splay, twist, and anisotropy moduli, respectively.
For PbTiO$_3$, the corresponding values (in the same units as $G_{ijkl}$) are
$K_{1}=G_{1122}+2G_{1212}=2.77$,
$K_{2}=2G_{1212}=2.77$, and 
$K_{\mathrm{a}}=G_{1111}-G_{1122}-2G_{1212}=0$.
The first two terms are rotationally invariant, while the third term reflects the underlying crystalline anisotropy and has no direct analogue in the classical Frank theory of nematics. Remarkably, this anisotropy contribution vanishes for PbTiO$_3$.

\subsection{Elastic energy of strained films}
The elastic energy $F_{u}$, expressed in terms of the strain tensor $u_{ij}$, in the general case takes the form  
\begin{equation}
    F_{u} = \tfrac{1}{2} C_{ijkl}\, u_{ij}u_{kl},
    \label{Fu}
\end{equation}
where $C_{ijkl}$ denotes the elastic stiffness tensor. In ferroelectrics, strain is typically induced by the polarization field through electrostrictive coupling. The condition of mechanical equilibrium imposes the constraint
\begin{equation}
   C_{ijkl}\,\partial_{j}\!\left(u_{kl}-Q_{klmn}P_{m}P_{n}\right)=0,
   \label{ElastEq}
\end{equation}
where $Q_{klmn}$ denotes the electrostrictive coefficients.

For epitaxial films clamped to a substrate, the in-plane strains are fixed as 
$u_{11}=u_{22}=u_m$ and $u_{12}=0$, while the out-of-plane components 
$u_{33}$, $u_{13}$, and $u_{23}$ are free to relax. Under these boundary 
conditions, the elastic energy can be minimized to eliminate the strain 
degrees of freedom. Following the approach of Pertsev and Tagantsev 
(P-T)~\cite{Pertsev1998}, the resulting effective free energy density 
retains the form of Eq.~(\ref{FP}), but with orthorhombic symmetry and 
renormalized coefficients of the second- and fourth-order terms.

The coefficients of the quadratic terms depend on both temperature $T$ and misfit strain $u_m$. This dependence determines the critical transition temperatures to the different ferroelectric phases as a function of $u_m$. For PbTiO$_3$, the renormalized values of these coefficients, expressed in units of $\mathrm{C^{-2}m^2N}$, are  
$a_{1,2}^*(T,u_m)  = a_1(T) - 11.0\,u_m \times 10^9$,  
and  
$a_3^*(T,u_m) = a_1(T) + 9.5\,u_m \times 10^9$,  
where $a_1(T)=3.8\times 10^5(T-479^\circ\mathrm{C})$ denotes the corresponding coefficient of the cubic, unstrained crystal.  
The coefficients of the quartic terms are renormalized as functions of the elastic parameters of the system. For PbTiO$_3$, expressed in units of $10^{-9}\,\mathrm{C^{-4}m^6N}$, they are:
 $a_{11}^* = a_{22}^* = 0.42$,
$a_{33}^*= 0.05$, 
$a_{13}^*= a_{23}^* = 0.45$,
$a_{12}^*= 0.73$.
The coefficients of the sixth-order terms remain non-renormalized and coincide with those of the unstrained cubic crystal: $a_{111}=0.26$, $a_{112}=0.61$, and $a_{123}=-3.71$ (in units of $10^{-9}\,\mathrm{C^{-6}m^{10}N}$), together with all equivalent cubic permutations.

\begin{figure}[ht!]
\centering  \includegraphics[width=0.5\textwidth]{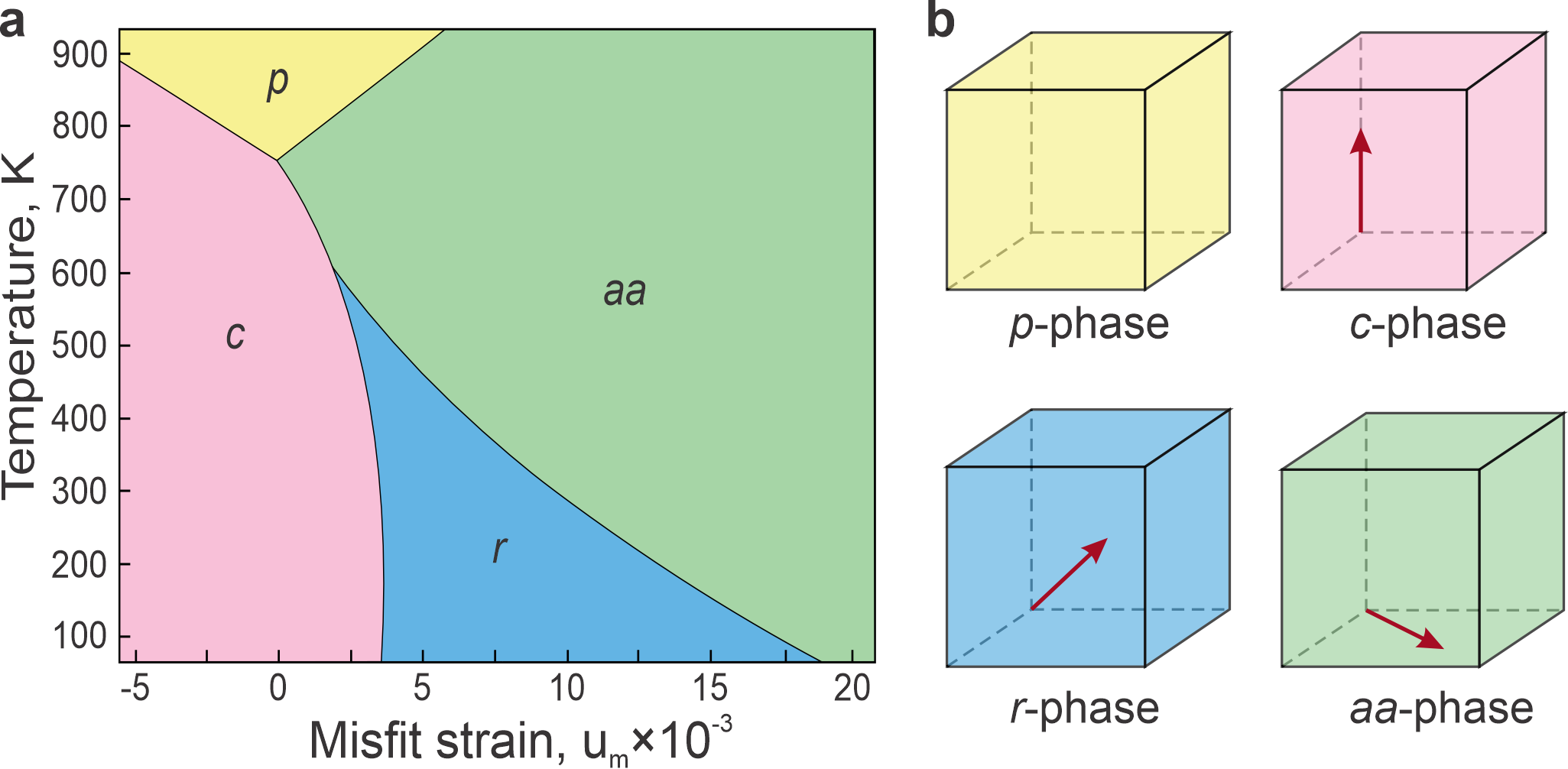}
    \caption{\textbf{Ferroelectric phases in strained PbTiO$_3$ films.}  
\textbf{a} Strain-temperature ($u_m$-$T$) phase diagram, as calculated in Ref.~\cite{Pertsev1998}.  
\textbf{b} Polarization configurations in the strained ferroelectric phases.}
    \label{FigPT}
\end{figure}

The strain-temperature $u_m$-$T$ phase diagram of strained PbTiO$_3$ films, 
derived from the renormalized P-T functional~\cite{Pertsev1998} is shown in Fig.~\ref{FigPT}a. 
Starting from the paraelectric $p$-phase, three distinct ferroelectric 
phases may emerge: the tetragonal $c$-phase 
$\mathbf{P}=(0,0,P_3)$, the orthorhombic $aa$-phase 
$\mathbf{P}=(P_1,P_1,0)$, and the monoclinic $r$-phase 
$\mathbf{P}=(P_1,P_1,P_3)$, as illustrated in Fig.~\ref{FigPT}b. 
These phases are distinguished by the orientation of the 
spontaneous polarization: in the $c$-phase, it is strictly 
out-of-plane; in the $aa$-phase, it is confined within the 
film plane; whereas in the $r$-phase, the polarization exhibits 
both in-plane and out-of-plane components.

The emergence of in-plane $aa$- and $r$-phases under tensile biaxial strain can be qualitatively understood as follows. The polarization state with two in-plane components allows both in-plane axes to contract via electrostriction, recovering nearly twice the elastic energy compared to the single-axis $a$-state that normally appears in freestanding bulk crystals. In practice, however, the system tends to split into alternating $a_{1}/a_{2}$ stripe domains~\cite{Pompe1993,Koukhar2001,Li2017}, which compensate the epitaxial strain even more effectively. Such multidomain relaxation lies beyond the P-T approach, which assumes uniform polarization states only.

\subsection{Electrostatic energy}
\label{SecElectrostatics}
The electrostatic contribution $F_{E}$ to the free energy density reads
\begin{equation}
F_{E} = \tfrac{1}{2}\,\varepsilon_0\varepsilon_b 
 \mathbf{E}^2  ,
\label{Fel}
\end{equation}
where the internal electric field $\mathbf{E}$ is dominated in confined ferroelectrics by the depolarizing field $\mathbf{E}_{\mathrm{dep}}$, produced by the bound charge 
$\rho_b = -\nabla \cdot \mathbf{P}$ and found from Poisson's equation for the electrostatic potential $\varphi$,  $\varepsilon_0\varepsilon_b\nabla^2\varphi=-\rho_b$ and $\mathbf{E}=-\nabla \varphi$. Here, $\varepsilon_b\simeq10$ is the background dielectric constant, associated with the non-polar atoms of the crystal lattice.

For a uniformly $z$-polarized PbTiO$_3$ film of thickness $2a_f$ with polarization vector $\mathbf{P}=(0,0,P_s)$ and spontaneous polarization $P_s \simeq 0.7\,\mathrm{C/m^{2}}$, the bound surface charges $\sigma_b=\pm P_s$ induce a homogeneous depolarizing field of magnitude
$
E_{\mathrm{dep}} = -{P_s}/{\varepsilon_0 \varepsilon_b}
\approx 7.8\times 10^{9} \,\mathrm{V/m}
$.
The corresponding electrostatic energy density is
$
F_{\mathrm{dep}} = \tfrac{1}{2}\varepsilon_0\varepsilon_b E_{\mathrm{dep}}^2
\approx 2.7\times 10^{9} \,\mathrm{J/m^{3}} .
$

In comparison, the condensation energy of the ferroelectric $c$-phase, estimated from the P-T energy density, is $F_P \approx -5.8\times 10^{7} \,\mathrm{J/m^{3}}$, nearly two orders of magnitude smaller. This strong difference indicates that the homogeneous state is electrostatically unstable. In order to suppress the large depolarization contribution, the system enforces an almost divergence-free condition for the polarization field,
\begin{equation}
\nabla \cdot \mathbf{P} \approx 0 .
\label{divP=0}
\end{equation}
This constraint removes bulk and surface bound charges and aligns polarization tangentially at free surfaces, thereby reducing the depolarizing field and giving rise to divergence-free polarization patterns such as vortices and bubbles, predicted in~\cite{Stephanovich2003,Naumov2004} and observed experimentally in PbTiO$_3$/SrTiO$_3$~\cite{Yadav2016,Yadav2019}.

These electrostatic considerations go beyond the original scope of the P-T approach~\cite{Pertsev1998,Pertsev1999}, which was formulated for strained monodomain configurations and does not explicitly include depolarization effects arising from surface bound charges. In the present work, we extend this framework by developing a consistent description of nonuniform polarization states within the soft-domain approach, which naturally incorporates domain formation with divergence-free polarization textures.

\section{Soft domains}
\label{SecSoftDomains}

\subsection{Approach}
In confined ferroelectrics, the full GLD  functional leads to a nonlinear system of coupled partial differential equations for the polarization field under elastic and electrostatic boundary conditions. Direct solutions require large-scale phase-field simulations or atomistic modeling, which, while accurate, are computationally demanding and often mask the underlying physical mechanisms.

A practical simplification is obtained by projecting the problem onto a reduced set of harmonic modes that automatically satisfy the boundary conditions and symmetries of thin films and superlattices. In this way, the infinite-dimensional variational problem formulated in the Hilbert space of polarization fields is reduced to an effective free energy functional depending only on a few modal amplitudes. The suggested in~\cite{Stephanovich2003,Stephanovich2005,Deguerville2005,Lukyanchuk2009,Kondovych2025} soft-domain approach complements numerical simulations by providing an exploratory and interpretative tool that guides the understanding and prediction of topological and structural states in confined ferroelectrics.

Within the GLD variational framework, including both uniform~(\ref{FP}) and gradient~(\ref{Fgrad}) contributions, the variation with respect to $\mathbf{P}$, 
\begin{equation}
{\delta \mathcal{F}}/{\delta \mathbf{P}}=0, 
\label{Variat}
\end{equation}
leads to a nonlinear system of coupled partial differential equations.
Near the ferroelectric transition, the small amplitude of the order parameter permits linearization, reducing the system to an anisotropic Laplace-Helmholtz equation for the polarization field,
\begin{equation}
A_{ij}(T)P_j - G_{ijkl}\,\partial_j\partial_l P_k = 0,
\label{GLDlin}
\end{equation}
with $A_{ij}(T)=\mathrm{diag}(a_1,a_2,a_3)$ and $G_{ijkl}$ the gradient coefficients.

The soft-domain approach consists in expanding the polarization in the eigen plane-wave modes of the linearized equation~(\ref{GLDlin}),
\begin{equation}
\mathbf{P}(\mathbf{r})=\sum_{\mathbf{k}} \mathbf{e}_j\,P_j(\mathbf{k})\,e^{i\mathbf{k}\cdot\mathbf{r}}+\mathrm{c.c.},
\label{Fourier}
\end{equation}
in which we enforce the electrostatic constraint~(\ref{divP=0}), leading to the condition 
\begin{equation}
 k_j P_j(\mathbf{k})=0,
\label{constraint}
\end{equation}
{\it i.e.}, each contributing mode is transverse to its wave vector $\mathbf{k}$.

Equations~(\ref{GLDlin}) reduce to an algebraic eigenvalue problem for the small Fourier amplitudes $P_j(\mathbf{k})$,    
\begin{equation}
\big[\,A_{ij}(T)+G_{ijmn}\,k_m k_n\,\big]\,P_j(\mathbf{k})=0.
\label{Eigen}
\end{equation}
Nontrivial solutions require  
$\det [A_{ij}(T)+G_{ijmn}k_mk_n]\,=\,0$.  
The realized soft mode corresponds to the maximal critical temperature of the solution of~(\ref{Eigen}), and the associated wave vector $\mathbf{k}_\ast$ sets the modulation periodicity.

In the spirit of Landau theory, we assume that close to the transition the spatial profile of the inhomogeneous structure, $\mathbf{P}(r)$, remains essentially unchanged, while its amplitude grows upon cooling. This assumption, originally justified in Ref.~\cite{Lukyanchuk2009}, underpins the soft-domain approach and allows one to capture the onset of modulated states without solving the full nonlinear problem.

Note that elastic couplings are nonlinear in $\mathbf{P}$ and negligible near $T_c$; at lower $T$, they become relevant ({\it e.g.}, $a_1/a_2$ twins in epitaxial films) and can be included effectively via the renormalization of quartic coefficients ({\it cf.}~P-T). The full elastic-driven texture analysis lies beyond our present scope.

\begin{figure*}[t!]
\centering  \includegraphics[width=0.9\textwidth]{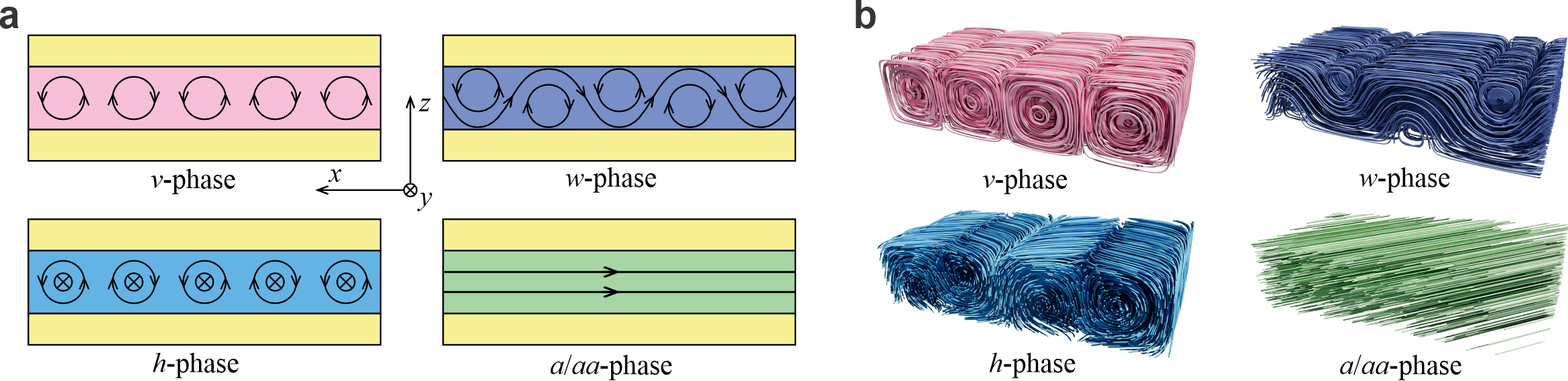}
\caption{\textbf{Topological polarization phases in thin ferroelctric films.} 
\textbf{a}  Morphological origin of emergent phases.
    \textbf{b}} 3D visualization of polarization streamlines.
    \label{FigPhases}
\end{figure*}

\subsection{Vortex states}
\label{SecVortexStates}
To describe inhomogeneous polarization states in ferroelectric thin films, we adopt the soft-domain approach originally introduced in Refs.~\cite{Stephanovich2003,Lukyanchuk2009}. The key idea is to replace the full solution of the nonlinear variational equations with a reduced representation that captures the essential topology of the polarization field, while satisfying the electrostatic constraint.

We consider a ferroelectric layer of thickness $2a_f$ along the $z$-axis hosting a periodic polarization domain structure along the $x$-axis with half-period $d$. The polarization distribution is expressed in the form of the lowest Fourier harmonics,
\begin{gather}
\mathbf{P}(\mathbf{r})=
\left( P_{a}\sin \frac{\pi z}{2a_{f}}\sin \frac{\pi x}{d}
+P_{1}\right)\mathbf{e}_{x} \notag \\
+P_{2}\,\mathbf{e}_{y}
+P_{3}\cos \frac{\pi z}{2a_{f}}\cos \frac{\pi x}{d}\,\mathbf{e}_{z}.
\label{ansatz}
\end{gather}
Here, $P_{1}$ and $P_{2}$ are the uniform in-plane components of polarization, while the vortex-like distortion of polarization in the $(x,z)$ plane is accounted for by the parameters $P_3$ and $P_a$. The sinusoidal form ensures smooth variations across the domain walls and automatically enforces tangential orientation of $\mathbf{P}$ at the film surfaces ($z=\pm a_f$), eliminating bound surface charges.

The electrostatic condition of vanishing divergence, $\nabla \cdot \mathbf{P} = 0$, imposes a relation between the modulation amplitude $P_a$ and the out-of-plane component $P_3$:
\begin{equation}
P_a = \gamma P_3,
\qquad
\gamma = \frac{d}{2a_f}.
\label{Pa}
\end{equation}
This condition guarantees the absence of bulk bound charges and therefore suppresses the depolarizing field. The entire vortex structure is then parameterized by only three amplitudes $(P_1, P_2, P_3)$ and the geometric parameter $\gamma$. The soft-domain ansatz (\ref{ansatz}) provides a natural parametrization for a broad family of polarization textures. This framework covers the various topological states that can be realized in ferroelectric thin films, depending on the amplitudes $(P_1,P_2,P_3)$. The basic representatives of these states are illustrated in Fig.~\ref{FigPhases}, summarized in Table~\ref{TablePhases}, and described in detail below.

\begin{table}[h!]
    \centering
    \begin{tabular}{lcl}
       \hline \hline
       Phase 
       & $(P_1 P_2 P_3)$  
       & Structure  \\
       \hline 
        Vortex $v$    & $(0\, 0\, P_3)$     & vortices aligned along [010] \\
        Helix $h$     & $(0 P_2 P_3)$    & [010]-vortices and uniform  $\mathbf{P}\parallel$\,[010] \\ 
        Wave w   & $(P_1 0 P_3)$  & [010]-vortices and uniform  $\mathbf{P}\parallel$\,[100]   \\
        Combined $wh$  & $(P_1 P_2 P_3)$   &   [010]-vortices and uniform  $\mathbf{P}\parallel$\,[a$_1$a$_2$0]  \\
        Uniform $a$   & $(P_1\, 0\, 0)$     & Uniform  $\mathbf{P}\parallel$\,[100] \\
        Uniform $aa$  & $(P_1 P_1\, 0)$   & Uniform  $\mathbf{P}\parallel$\,[110] \\
       \hline \hline
    \end{tabular}
    \caption{Polarization states in ferroelectric thin films obtained within the soft-domain ansatz.}
    \label{TablePhases}
\end{table}

\textit{Vortex phase} $v$, $(0,0,P_3)$.
The polarization lines form closed loops in the $(x,z)$ plane, yielding a periodic array of vortices and antivortices. This state minimizes the depolarizing energy by enforcing tangential polarization at the film surfaces.

\textit{Helix phase} $h$, $(0,P_2,P_3)$.
A uniform in-plane component $P_2$ is superimposed on the vortex texture. This converts the alternating vortex-antivortex pattern into a chiral arrangement where all vortex cores carry the same axial polarization, resulting in a helical state.

\textit{Wave phase} $w$, $(P_1,0,P_3)$.
A uniform component $P_1$ along $x$ shifts neighboring vortices vertically in opposite directions, creating a polarization wave with alternating modulation along the [100] axis.

\textit{Uniform in-plane phases ($a$, $aa$)}.
For $(P_1,0,0)$, the polarization is uniformly aligned along $x$ (phase $a$). For $(P_1,P_1,0)$, polarization lies uniformly in the film plane along [110] (phase $aa$). These phases correspond to classical strain-stabilized ferroelectric states in epitaxial films.

 Combinations of these basic states can also occur. For example, the superposition of the wave ($w$) and helix ($h$) states gives rise to the \textit{wave-helix phase} $wh$, defined by $(P_1,P_2,P_3)$. When both in-plane components, typically with $P_1 \approx P_2$, are present, the system develops a polarization wave structure,
with polarization streamlines propagating approximately along the [110] direction.

We also note that the principal modulations may orient along diagonal directions in the film plane. In such cases, the corresponding [110]-vortex states are denoted with a prime, for instance, $v'$-, $w'$-,  $h'$-phases.

\subsection{Effective energy}

To determine equilibrium states within the soft-domain approach, we substitute the inhomogeneous polarization ansatz~(\ref{ansatz}) into the uniform polarization functional~(\ref{FP}) for a strained epitaxial film with P-T strain-renormalized coefficients $a_i^*$ and $a_{ij}^*$ and a gradient energy term~(\ref{Frank}). Averaging over the domain-cell volume yields a polynomial free energy expansion in a reduced set of modal amplitudes $P_1$, $P_2$, $P_3$, and the structural parameter $\gamma$~(\ref{Pa}), which fixes the domain period.

The following averages of powers of sine and cosine functions are used to calculate the soft domain energy,
\begin{gather}
\langle \cos^{2} \rangle = \langle \sin^{2} \rangle = \tfrac{1}{2}, \quad
\langle \cos^{4} \rangle = \langle \sin^{4} \rangle = \tfrac{3}{8}, 
\notag \\
\langle \cos^{6} \rangle = \langle \sin^{6} \rangle = \tfrac{5}{16}, \quad
\langle \cos^{2} \sin^{2} \rangle = \tfrac{1}{8}, 
\notag \\
\langle \cos^{2} \sin^{4} \rangle = \langle \cos^{4} \sin^{2} \rangle = \tfrac{1}{16}. 
\label{sinaverage}
\end{gather}

The resulting effective free energy functional has the same algebraic form as Eq.~(\ref{FP}), with coefficients expressed through the P-T coefficients of the strained film. The explicit expressions for the expansion coefficients $\alpha$ are summarized in Table~\ref{TableSoft}.
The last line corresponds to the gradient contribution~(\ref{Frank}) rewritten in a simple form, $F_{\mathrm{grad}}=(\beta_2+\beta_a)P_3^2$.
\begin{table*}[!t]
\centering
\begin{tabular}{ll}
\hline \hline 
$\alpha _{1}=a^*_{1},\ \ \ $ $\alpha _{2}=a^*_{1},$ & $\alpha _{111}=a_{111}$, \ \ \  $\alpha
_{222}=a_{111},$ \\ 
$\alpha _{3}=\frac{1}{4}\gamma ^{2}a^*_{1}+\frac{1}{4}a^*_{3},$ & $\alpha _{333}=%
\frac{25}{256}a_{333}+\frac{1}{256}\gamma ^{2}a_{133}+\frac{1}{256}\gamma
^{4}a_{113}+\frac{25}{256}\gamma ^{6}a_{111},$ \\ 
$\alpha _{11}=a_{11},$ & $\alpha _{112}=\alpha _{112}$, \ \ \  $a_{122}=a_{112},$
\\ 
$\alpha _{22}=a^*_{11},$ & $\alpha _{113}=\frac{1}{4}a_{113}+\frac{15}{4}%
\gamma ^{2}a_{111},$ \\ 
$\alpha _{33}=\frac{9}{64}\gamma ^{4}a^*_{11}+\frac{9}{64}a^*_{33}+\frac{1}{64}%
\gamma ^{2}a^*_{13},$ & $\alpha _{223}=\frac{1}{4}a_{113}+\frac{1}{4}\gamma
^{2}a_{112},$ \\ 
$\alpha _{12}=a^*_{12},$ & $\alpha _{133}=\frac{9}{64}a_{133}+\frac{6}{64}%
\gamma ^{2}a_{113}+\frac{135}{64}\gamma ^{4}a_{111},$ \\ 
$\alpha _{13}=\frac{3}{2}\gamma ^{2}a^*_{11}+\frac{1}{4}a^*_{13},$ & $\alpha
_{233}=\frac{9}{64}a_{133}+\frac{1}{64}\gamma ^{2}a_{123}+\frac{9}{64}\gamma
^{4}a_{112},$ \\ 
$\alpha _{23}=\frac{1}{4}a^*_{13}+\frac{1}{4}\gamma ^{2}a^*_{12},$ & $\alpha
_{123}=\frac{1}{4}a_{123}+\frac{3}{2}\gamma ^{2}a_{112}.$ \\ [4pt]
$\beta_2=\frac{\pi ^{2}}{8}\frac{K_{2}}{( 2a^*_{f}) ^{2}}%
( \gamma +\gamma ^{-1}) ^{2}$,
&
$\, \, \beta_a=\frac{\pi ^{2}}{8}\frac{2K_{a}}{%
( 2a_{f}) ^{2}}$\\ [4pt]
\hline \hline 
\end{tabular}
\caption{Soft-domain functional coefficients.}
\label{TableSoft}
\end{table*}
\renewcommand{\arraystretch}{1.0}
\setlength{\tabcolsep}{6pt}


It should be emphasized that here, $P_1$, $P_2$, and $P_3$ are not the components of a uniform polarization but rather amplitudes parametrizing the inhomogeneous texture~(\ref{ansatz}) inside a soft domain.

To describe configurations in which the vortices are aligned along the in-plane diagonals, {\it i.e.}, with their cores oriented along the $[110]$ directions of the cubic lattice, we perform a $45^\circ$ rotation of both the in-plane polarization and the coordinate frame about the $z$ axis. Denoting the rotated variables with primes, the transformation reads
\begin{gather}
    P_1=\tfrac{1}{\sqrt{2}}\,(P'_1+P'_2),\quad 
    P_2=\tfrac{1}{\sqrt{2}}\,(P'_1-P'_2),\quad 
    P_3=P'_3, \notag \\
    x=\tfrac{1}{\sqrt{2}}\,(x'+y'),\quad 
    y=\tfrac{1}{\sqrt{2}}\,(x'-y'),\quad 
    z=z'.
  \label{rot45}
\end{gather}
Applying this $45^\circ$ rotation to the P-T functional, we obtain an analogous free energy expansion with modified coefficients, listed in Table~\ref{Tab45rot}. Importantly, after rotation, the coefficients of the sixth-order invariants no longer satisfy the cubic symmetry relations of the original P-T functional. Note that the  coefficients in the gradient energy~(\ref{Frank}) remain unchanged because the first two terms are rotationally invariant, while the coefficient $K_a$ of the third, anisotropic, contribution vanishes for PbTiO$_3$. 

The subsequent minimization procedure is then carried out by averaging according to Eq.~(\ref{sinaverage}), which yields an effective soft-domain functional with coefficients summarized in Table~\ref{TableSoft}. In this effective description, the original P-T coefficients are systematically replaced by their rotated counterparts from Table~\ref{Tab45rot}. 

This procedure allows us to evaluate the energies of phases in which the vortices are aligned along the in-plane diagonals ($v'$, $h'$, $w'$) and to compare them with the corresponding vortex phases oriented along the principal crystallographic axes ($v$, $h$, $w$). By identifying the states of lowest free energy, we can determine the stability regions of the diagonal and axial vortex configurations and construct the phase diagrams of the system, which are discussed in the following section.

\section{Topological Polarization States}
\label{SecPolarizationStates}

\begin{figure*}[ht!]
\centering  \includegraphics[width=1.0\textwidth]{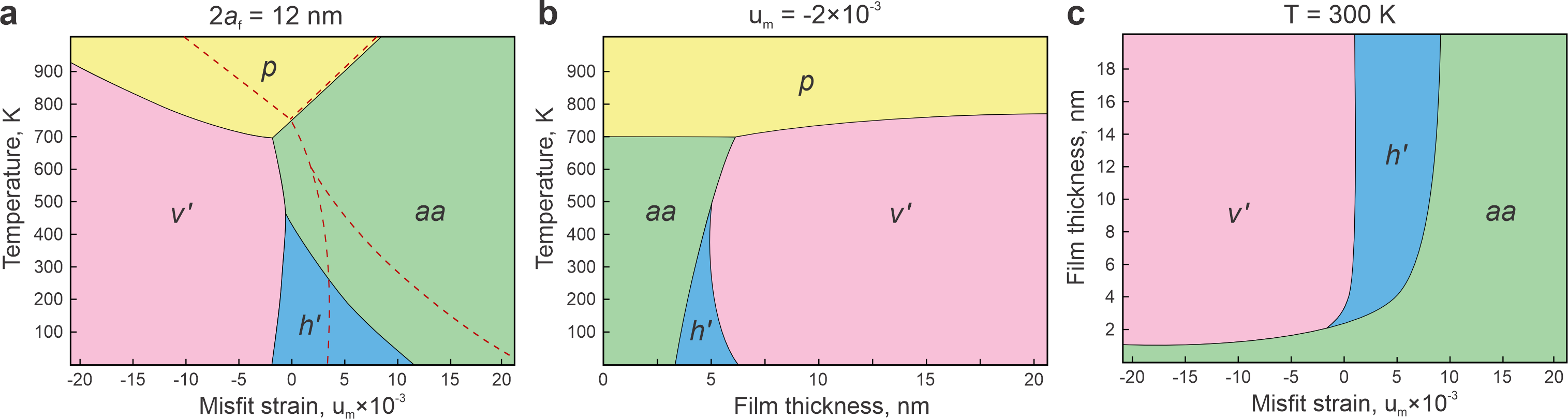}
    \caption{\textbf{Soft-domain ferroelectric phases in strained PbTiO$_3$ films.}  
\textbf{a} Strain-temperature ($u_m$-$T$) phase diagram for a film of thickness $2a_f=12$~nm. The dashed red lines present the original P-T phase diagram, given in Fig.~\ref{FigPT}.  
\textbf{b} Film thickness-temperature ($2a_f$-$T$) phase diagram in a film under misfit strain $u_m=-2 \times 10^{-3}$. 
\textbf{c} Strain-film thickness ($u_m$-$2a_f$) phase diagram at room temperature $T=300$~K.}
    \label{FigPhaseDi}
\end{figure*}

\begin{figure*}[ht]
    \centering
    \includegraphics[width=\textwidth]{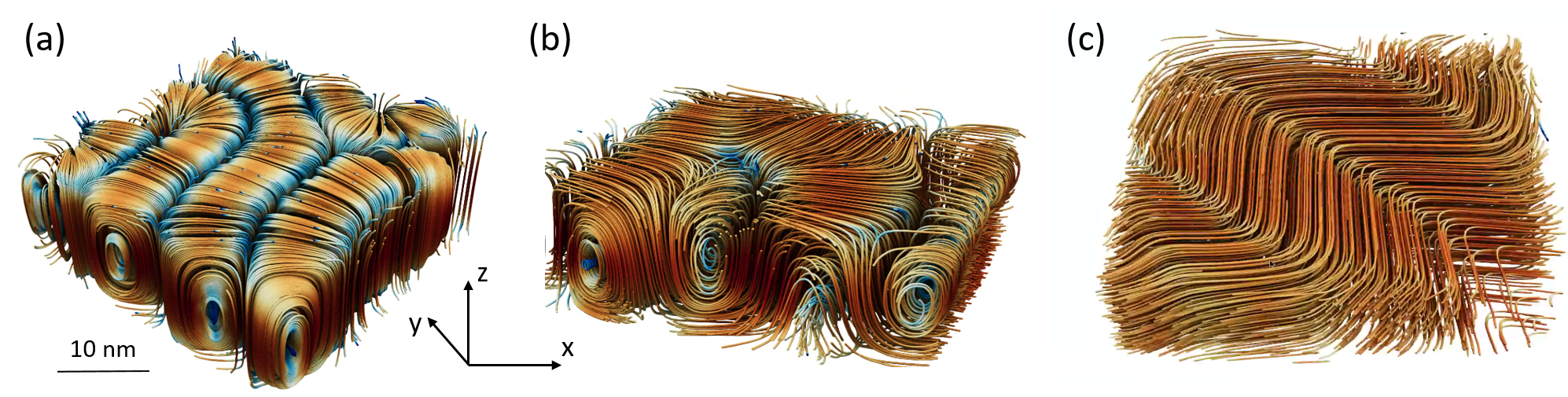}
    \caption{\textbf{Phase-field modeling of topological states in strained films of PbTiO$_3$}. 
\textbf{a}  Polarization texture in a rotated vortex $v'$-phase in the PbTiO$_3$ layer of thickness $2a_f=12$\,nm at $T=300$\,K and $u_m=-0.01$. 
\textbf{b}  Wave-helix $wh$-phase at $T=300$\,K and $u_m\simeq 0$.
\textbf{c}  $a_1/a_2$ domains at $T=300$\,K and $u_m=0.01$.
    }
    \label{FigSimul}
\end{figure*}

\subsection{Phase diagrams}

Applying the soft-domain variational minimization developed in the previous section, and using the specific GLD and elastic coefficients of PbTiO$_3$, we now construct phase diagrams for strained PbTiO$_3$ films that summarize the stability regions of the different polarization states. By analyzing representative cross-sections in the parameter space of misfit strain $u_m$, temperature $T$, and film thickness $2a_f$, we identify the regimes where uniform and nonuniform states emerge and compete.

Two nonuniform polarization states, the rotated vortex $v'$-phase and the rotated helix $h'$-phase, emerge in a strained film, as shown in Fig.~\ref{FigPhaseDi}a. The $v'$-phase stabilizes at negative misfit strain and can be regarded as the analogue of the uniform $c$-phase, while the $h'$-phase corresponds to the uniform $r$-phase. Tensile misfit strain stabilizes the uniform $aa$-phase, in close analogy to the P-T case. The transition temperature from the paraelectric phase into the $v'$-phase is reduced compared to the $c$-phase transition (without accounting for depolarization effects), since the formation of vortices requires additional energy.

Taking into account the energy dependence on the structural parameter $\gamma$ specified in~(\ref{Pa}), we present film thickness-temperature $2a_f$-$T$ phase diagram for a PbTiO$_3$ film under the compressive misfit strain $u_m=-2 \times 10^{-3}$, see Fig.~\ref{FigPhaseDi}b. We observe that the rotated vortex $v'$-phase emerges in the wide range of film thicknesses, $2a_f \gtrsim 5$~nm. At film thicknesses of around 5~nm, the transient area of the rotated helix $h'$-phase is observed.
The uniform $aa$-phase is favorable in ultrathin films $\lesssim 5$~nm; however, it is expected that polar phases are strongly suppressed at small film thicknesses.

To complete the overview of phase stabilization conditions, we present the misfit strain-film thickness $u_m$-$2a_f$ phase diagram at room temperature, $T=300$~K, shown in Fig.~\ref{FigPhaseDi}c. We observe that the compressive misfit strain stabilizes the rotated vortex $v'$-phase, while the small tensile strain favors the emergence of $h'$-phase. At larger tensile strains, as well as in ultrathin PTO films, the uniform $aa$-phase emerges. 

Notably, in the transient region between uniform and nonuniform states, the energy of the $h'$-phase is comparable to that of the $h$, $w$, $hw$, and $w'$ topological phases. These states can all be interpreted as different superpositions of the uniform $a$- or $aa$-polarization with vortex arrays, distinguished only by their relative orientation and the way the vortex component is combined with the in-plane order. Due to the small energy difference between such configurations, ferroelectric ordering becomes highly sensitive to the precise values of geometric parameters, which may lead to the coexistence of metastable clusters associated with different textures. This picture is further supported by phase-field simulations, which reveal the formation of such complex mixed patterns.

\setlength{\tabcolsep}{12pt}
\renewcommand{\arraystretch}{1.3}
\begin{table}[b]
\centering
\label{TableSoftRot}
\begin{tabular}{ll}
\hline \hline 
$a_{1}^{*\prime }=a_{1}^*$, \  $a_{3}^{*\prime }=a_{3}^*$, & $a_{111}^{\prime }=\frac{1%
}{4}a_{111}+\frac{1}{4}a_{112}$, \\ 
$a_{11}^{*\prime }=\frac{1}{2}a_{11}^*+\frac{1}{4}a_{12}^*$, & $a_{333}^{\prime
}=a_{333}$, $\ a_{133}^{\prime }=a_{133}$, \\ 
$a_{33}^{*\prime }=a_{33}^*$, & $a_{112}^{\prime }=\frac{15}{4}a_{111}-\frac{1}{4%
}a_{112}$, \\ 
$a_{12}^{*\prime }=3a_{11}^*-\frac{1}{2}a_{12}^*$, & $a_{113}^{\prime }=\frac{1}{2}%
a_{113}+\frac{1}{4}a_{123}$, \\ 
$a_{13}^{*\prime }=a_{13}^*$, & $a_{123}^{\prime }=3a_{113}-\frac{1}{2}a_{123}$  \\ 
\hline \hline 
\end{tabular}
\caption{P-T coefficients after $45^\circ$ in-plane rotation.}
\label{Tab45rot}
\end{table}
\renewcommand{\arraystretch}{1.3}
\setlength{\tabcolsep}{6pt}

\subsection{Phase-field modeling}
We carried out full phase-field simulations of strained PbTiO$_3$ films to establish a framework that demonstrates the consistency of the soft-domain method and places it within a broader modeling context.

In the phase-field approach, the minima of the free energy functional~(\ref{FunctGLD}) are obtained by solving the nonlinear relaxation equations derived from the variational condition~(\ref{Variat}). For numerical implementation, the functional~(\ref{FunctGLD}) includes the uniform contribution~(\ref{FP}), the gradient term~(\ref{Fgrad}), the elastic energy~(\ref{Fu}), and the electrostatic energy~(\ref{Fel}). 

To simulate the formation of inhomogeneous polar structures in ferroelectric layers, we considered the periodically repeated unit of a model superlattice composed of PbTiO$_3$ combined with SrTiO$_3$ or with another dielectric of comparable permittivity, with individual layer thicknesses of 12nm\,/12nm. Three representative samples were selected and subjected to global epitaxial misfit strains of $u_m=-0.01$, $0$, and $+0.01$, respectively, at room temperature. Such simulations refine the average description of deformation within the P-T approach by providing access to the full spatial distribution of strain.

The simulations were performed with the open-source FEniCS framework~\cite{LoggMardalEtAl2012a}. Three-dimensional finite element meshes partitioned into unstructured tetrahedrons were generated with the help of the 3D mesh generator \emph{gmsh}~\cite{Geuzaine2009}. The unknown variables \textbf{P}, $\varphi$, and $u_{ij}$, for the case if the elastic contribution is included, were approximated by piecewise linear polynomials. The BDF2 time stepper with adaptive step size~\cite{Janelli2006} was used to approximate the time derivative.

At the first step of each simulation, the initial paraelectric state was set as a random distribution of vector $\mathbf{P}$ components in the range of $-10^{-6}$ to $10^{-6}$~Cm$^{-2}$. The Newton method with line search was employed to solve the non-linear system arising from equation (\ref{Variat}). The linear systems on each nonlinear iteration and systems defined by Poisson's and linear elasticity (\ref{ElastEq}) equations were solved using the generalized minimal residual method with restart\,\cite{petsc-user-ref,petsc-web-page}.
The specific parameters and coefficients used in the phase-field simulations were taken from Refs.\,\cite{Lukyanchuk2024,Lukyanchuk2025}, where they were compiled from the original works with corresponding references.

The polarization distribution in three representative samples is illustrated in Fig.~\ref{FigSimul}. As anticipated in Sec.~\ref{SecElectrostatics}, the polarization streamlines align tangentially to the film surfaces, which strongly suppresses depolarizing effects, and the bound charges $\rho_b = -\nabla \cdot \mathbf{P}$ are found to be vanishingly small. In addition, quantitative analysis shows that the polarization profile across the film follows a smooth sinusoidal-like variation, in full agreement with the soft-domain description.

\begin{figure*}[!th]
\centering
\includegraphics[width=0.9\textwidth]{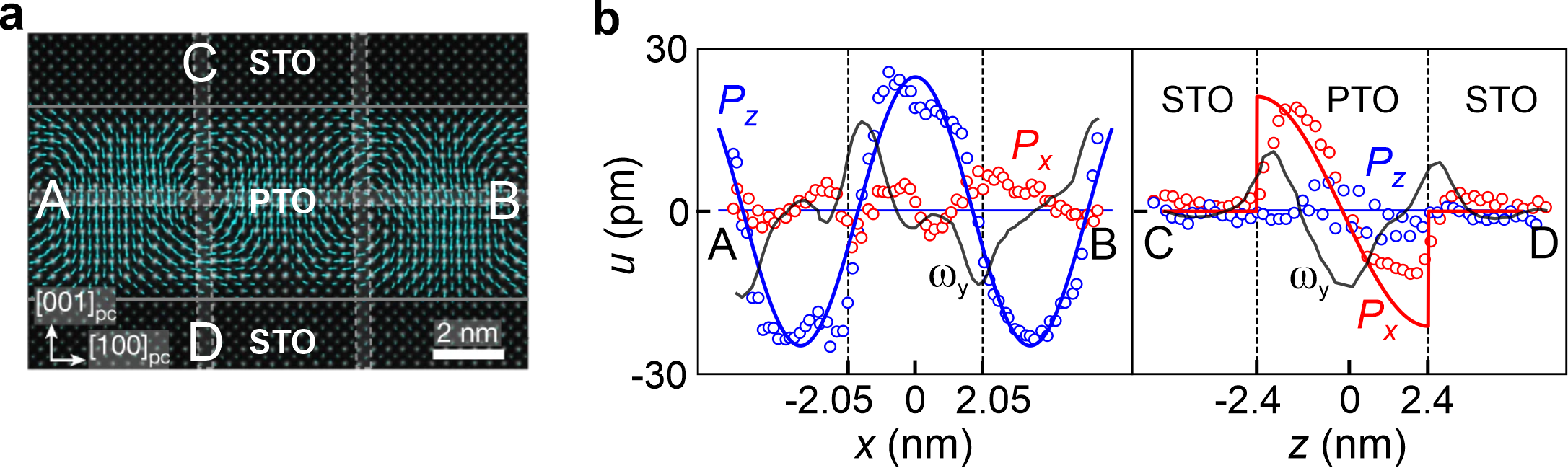}
\caption{\textbf{Soft-domain textures in ferroelectric heterostructures}.
\textbf{a} Polarization texture in a PbTiO$_3$ (PTO) layer of a PbTiO$_3$/SrTiO$_3$ superlattice.
\textbf{b} Experimental profiles of $u$, $P_x$, and $P_z$ along AB and CD (blue/red circles) compared with the soft-domain model~(\ref{ansatz}) (solid lines).
Adapted from Refs.~\cite{Yadav2019,Lukyanchuk2024}.}
\label{FigVF}
\end{figure*}

Figure~\ref{FigSimul}a shows the vortex phase stabilized under compressive misfit strain $u_m=-0.01$. The polarization streamlines form a vortex array structure aligned approximately along the in-plane diagonal, corresponding to the $v'$-phase in the phase diagrams of Fig.~\ref{FigPhaseDi}. Notably, disclination-type topological defects disrupt the regular arrangement of the vortex lattice.

The sample shown in Fig.~\ref{FigSimul}b, calculated at nearly relaxed misfit strain $u_m\simeq 0$, exhibits a more complex vortex pattern. Instead of forming a regular vortex array corresponding to the helical $h^\prime$-phase predicted in the phase diagrams of Fig.~\ref{FigPhaseDi}, the vortices arrange into a labyrinthine structure with signatures of several competing states. A clear helical bias along the axial direction is observed, characteristic of the $h$- and $h^\prime$-phases, while the vertical displacement of vortices is reminiscent of the $w$-type modulation. This irregular mixture suggests a configuration closer to a distorted $wh$-phase. However, it is likely that in this transient regime all intermediate vortex states, as well as differently oriented uniform $a$- or $aa$-polarization components discussed in Sec.~\ref{SecVortexStates}, correspond to metastable states with almost degenerate energies. We also note that their formation is influenced by local inhomogeneities of elastic strain, an effect not explicitly treated within the soft-domain approach.

The sample in Fig.~\ref{FigSimul}c, calculated at tensile misfit strain $u_m=+0.01$, shows a parquet-like arrangement of two domain types with polarization oriented along the [100] and [010] directions. This so-called $a_1/a_2$ domain structure is well-established in strained ferroelectric films~\cite{Pompe1993,Koukhar2001,Li2017} and represents the optimal configuration to minimize the inhomogeneous part of the elastic energy, which, as already noted, is not fully captured by the soft-domain approach. When averaged on length scales larger than the domain period, however, this state effectively reproduces the uniform $aa$-phase shown in the phase diagrams of Fig.~\ref{FigPhaseDi}.

 A detailed comparison of the results from the soft-domain approach and phase-field modeling with experimental observations is presented in the following section.

\section{Challenges and Perspectives}
\label{SecChallenges}

\subsection{Experiment vs modeling}

Early studies of ferroelectric nanostructures began with the prediction of classical Landau-Kittel-type domains, which already indicated that uniform polarization states are unstable against depolarizing fields\,\cite{Bratkovsky2000,Stephanovich2003,Deguerville2005}. Subsequent atomistic simulations~\cite{Naumov2004} showed that these stripe-like domains can evolve into flux-closure polarization vortices, representing a natural continuation of the same tendency to minimize depolarizing fields. The application of advanced experimental techniques then confirmed that the system is far more complex than a simple domain pattern, revealing a rich topological landscape of polarization vortices and more complex polarization swirl structures~\cite{Junquera2023,Lukyanchuk2025}.

A topological classification of polarization textures is constructed based on the nearly divergence-free character of the polarization field\,\cite{Lukyanchuk2024}. The soft-domain method, originally introduced in\,\cite{Stephanovich2003,Deguerville2005,Lukyanchuk2009}, is further developed in this work to complement the topological description by providing a morphological classification of vortex-type textures in ferroelectric films and heterostructures. Within this framework, the broad variety of observed states is reduced to a simple ansatz~(\ref{ansatz}), parametrized by a small set of variational parameters that determine the structural features of the polarization field. This reduction makes it possible to relate distinct textures within a unified hierarchy and to analyze the polarization structures in an intuitively clear quantitative way.

The experimental evidence for both the topological approach and the soft-domain method is illustrated in Fig.~\ref{FigVF}. In particular, Fig.~\ref{FigVF}a shows a vortex state in PbTiO$_3$/SrTiO$_3$ superlattices, where a nearly divergence-free polarization texture forms and the streamlines align tangentially to the film surfaces, thereby strongly suppressing bound charges. Figure~\ref{FigVF}b demonstrates the remarkable agreement between experimental polarization profiles and the sinusoidal fit provided by the soft-domain ansatz~(\ref{ansatz}).

Using soft-domain morphological classification, many experimentally observed polarization textures, which at first glance appeared as separate and unrelated findings, can  be consistently understood within a unified framework.
The predicted vortex states in Sec.~\ref{SecVortexStates}  have been observed in PbTiO$_3$/SrTiO$_3$ superlattices under different experimental conditions. The vortex $v$-phase was reported in Refs.\,\cite{Yadav2016,Yadav2019}, the helix $h$-phase in Ref.\,\cite{Shafer2018}, the wave $w$-phase in Ref.\,\cite{Gong2021}, and the combined wave-helix $wh$-phase in Ref.\,\cite{Behera2022}. 

Notably, beyond striped vortex arrays, experiments have also revealed the polarization segmentation into periodic bubble structures, skyrmion-like bubbles\,\cite{Das2019} and meron-like bubbles \,\cite{Wang2020,Shao2023}. Such configurations can likewise be described within the soft-domain framework, extending the plane-wave parametrization~(\ref{Fourier}) beyond a single-$\mathbf{k}$  wave to a superposition of several plane waves, typically forming a regular star of wave vectors $\{\mathbf{k}_i\}$ with equal amplitudes. For instance, a threefold star with wave vectors separated by $120^\circ$ naturally generates a hexagonal lattice of polarization bubbles, consistent with the experimentally observed skyrmion-bubble-like states.

\subsection{From qualitative to quantitative}

In the previous section, we compared the ansatz~\ref{ansatz} within the soft-domain framework with experimental observations, showing that the method captures the principal features of discovered topological states and provides a systematic morphological classification of polarization textures. Yet the approach goes significantly further, as it also enables a quantitative exploration of these states by means of relatively simple analytical tools, achieved through a reduction to a small set of variational parameters, without requiring extensive computational resources. This makes it possible to evaluate the free energy of different configurations and to construct phase diagrams, a task that remains a major challenge for numerical modeling strategies. As an example, we constructed phase diagrams of the states realized in strained PbTiO$_3$ films as functions of the principal control parameters of the system, namely film thickness, temperature, and epitaxial misfit strain.

However, when analyzing experimental observations and numerical modeling, the situation is far from straightforward. In practice, it is difficult to obtain systematic data across a broad range of parameters: experimental conditions are constrained by growth techniques and measurement resolution, while numerical phase-field simulations demand substantial computational resources and are often affected by finite-size effects. Despite this complexity, numerous studies devoted to constructing phase diagrams using various simulation techniques have identified stable and metastable nonuniform polarization states under varying boundary conditions and external constraints.

Atomic-scale mapping has revealed periodic dipole waves in PbTiO$_3$-based oxides, forming stripe-like oscillatory polarization states that replace uniform domains within certain stability ranges~\cite{Sheng2008}. Phase-field simulations of PbTiO$_3$ thin films under anisotropic misfit strain demonstrated transitions between $c$-domains and mixed $a$/$c$-domain patterns, with their stability strongly influenced by strain anisotropy~\cite{Hong2017}. In ferroelectric superlattices, vortex arrays and skyrmion-like bubble structures were predicted as equilibrium states stabilized by the interplay of electrostatic and elastic energies~\cite{Gong2021}, while further work on PbTiO$_3$/SrTiO$_3$ systems mapped the conditions for vortex lattices, flux-closure domains, and uniform ferroelectric states depending on strain and periodicity~\cite{Dai2023}. For (110)-oriented PbTiO$_3$ films, strain-strain diagrams revealed unconventional ferroelastic configurations, distinct from the usual stripe or vortex structures of (001)-oriented systems~\cite{Li2024}. Temperature-dependent studies further indicated switching between vortex lattices, skyrmion bubbles, and uniform ferroelectric phases, pointing to thermally induced multi-state behavior~\cite{Tong2025}. At a more conceptual level, such diagrams have been described as collective landscapes of dipolar order, where ordered vortex arrays, bubble lattices, and irregular ``murmurations'' of dipoles may coexist and transform under external control~\cite{Gregg2025}.  

In reality, experimental evidence, atomistic modeling, and phase-field simulations indicate that perfectly ordered vortex states are rarely stabilized in ferroelectric films. This is especially true in the intermediate regime of small misfit strains, corresponding to the $h'$ region of the phase diagrams in Fig.~\ref{FigPhaseDi}, where several competing textures with different vortex-like and uniform in-plane polarization components possess nearly degenerate energies.
 Consequently, the emergence of labyrinthine configurations with mixed rotational senses and variably oriented uniform components should be regarded as an intrinsic feature of the system. In such structures, observed experimentally~\cite{Nahas2020,Nahas2020b} and reproduced in our simulations (Fig.~\ref{FigSimul}b),  the topological analysis establishes the coexistence of distinct topological polarization structural units~\cite{Lukyanchuk2024,Lukyanchuk2025}, most notably vortices and hopfions, which shape the intricate labyrinthine textures. 
 Within this context, the soft-domain method provides a powerful framework: it predicts the parameter ranges where such transitional states occur and clarifies their morphology, while acknowledging that the exact microscopic realization remains highly sensitive to growth conditions and to the thermodynamic history of the sample.

\section{Conclusion}
\label{SecConclusion}
In this work we have explored the rich variety of topological polarization states that emerge in strained ferroelectric films. These states, ranging from simple vortex arrays to complex labyrinthine patterns, reflect the fundamental tendency of the system to minimize depolarizing fields and elastic energy under confinement. To analyze their origin and organization, we applied the soft-domain method, which provides a compact variational framework capable of reducing complex polarization morphologies to a small set of parameters. This reduction enables a transparent classification of the observed states, quantitative evaluation of their stability, and rapid construction of phase diagrams. While not a substitute for atomistic or phase-field simulations, the soft-domain approach complements them by clarifying the underlying mechanisms and guiding more detailed modeling. Looking ahead, extensions of the method to anisotropic strain conditions, flexoelectric couplings, and multilayer architectures promise to broaden its relevance for both fundamental studies and the design of functional ferroelectric nanostructures.

\section*{Acknowledgments}

This research was financially supported by the European Union action H2020-MSCA-SE-3D-TOPO, project number 101236483.
A.R. acknowledges the Slovenian Research Agency support (P1-0125). S.K. acknowledges the support from the Philipp Schwartz Initiative of the Alexander von Humboldt Foundation.

We acknowledge the granted access to high-performance computing resources of ``Plateforme MatriCS'' within University of Picardie Jules Verne,  co-financed by the European Regional Development Fund (FEDER) and the Hauts-De-France Regional Council.


%

\end{document}